# Partial Complementary Energy Densities, Their Variational Principles and Applications in Elasticity

Jiashi Yang (jyang1@unl.edu)
Department of Mechanical and Materials Engineering
University of Nebraska-Lincoln, Lincoln, NE 68588-0526, USA

**Abstract**

Partial complementary energy densities are introduced through partial Legendre transforms from the strain energy density of linear elasticity. They have mixed components of the strain and stress tensors. Mixed variational principles based on these energy densities are presented. It is shown that these variational principles are useful in the derivation of two- and one-dimensional theories of elastic plates and rods.

## 1. Introduction

In the theory of elasticity it is well known that there are two energy density functions, the strain energy and the complementary energy densities, and there are various variational principles associated with these two energy densities [1]. The strain energy density is a function of the six strain components, while the complementary energy density depends on the six stress components. It is also well known that the two energy densities are related by Legendre transforms which change the independent variables of one energy density, e.g., the strain components, to stress components or vice versa.

As to be seen below, the Legendre transforms may be partial, i.e., replacing some of the strain components in the strain energy density by the corresponding stress components. This results in partial complementary energy densities with six mixed strain and/or stress components. Some variational principles based on these partial complimentary energy densities are also given below. It is shown that the partial complementary energy densities and the corresponding variational principles are useful in the development of two- and one-dimensional theories of elastic plates and rods.

## 2. Strain and Complementary Energy Densities and Their Variational Principles in Elasticity

For convenience, a brief summary is given in this section for the equations of linear elasticity, the strain and complementary stress energy densities and their variational principles in the notation of [2] which is commonly used for anisotropic materials. The Cartesian tensor notation is used. We consider elastostatics only. The equilibrium equation is

$$T_{ji,j} + f_i = 0, \quad (1)$$

where **T** is the stress tensor. **f** is the body force per unit volume. Constitutive relations are given by a strain energy density function

$$U(\mathbf{S}) = \frac{1}{2} c_{ijkl} S_{ij} S_{kl} \quad (2)$$

through

$$T_{ij} = \frac{\partial U}{\partial S_{ij}} = c_{ijkl} S_{kl}, \tag{3}$$

where the strain tensor, **S**, is related to the displacement vector, **u**, by

$$S_{ij} = (u_{j,i} + u_{i,j})/2. \tag{4}$$

$c_{ijkl}$ is the elastic stiffness. (3) can be inverted for strain in terms of stress, i.e.,

$$S_{ij} = s_{ijkl} T_{kl}, \tag{5}$$

where $s_{ijkl}$ is the elastic compliance. (5) can also be obtained through a complementary energy density

$$U^*(\mathbf{T}) = \frac{1}{2} s_{ijkl} T_{ij} T_{kl} \tag{6}$$

through

$$S_{ij} = \frac{\partial U^*}{\partial T_{ij}} = s_{ijkl} T_{kl}. \tag{7}$$

The complementary energy density is related to the strain energy density through the Legendre transform:

$$U^*(\mathbf{T}) = T_{ij} S_{ij} - U(\mathbf{S}), \tag{8}$$

where the strain tensor on the right-hand side is viewed as a function of the stress tensor through (5).

An abbreviated, indicial notation is often used in which a pair of Cartesian tensor indices ranging over the integers 1, 2 and 3 is replaced by one index ranging over the integers 1, 2, 3, 4, 5 and 6 according to

| $ij$ or $kl$: | 11 | 22 | 33 | 23, 32 | 31, 13 | 12, 21 |
|---|---|---|---|---|---|---|
| $p$ or $q$: | 1 | 2 | 3 | 4 | 5 | 6 |

In this notation, for the stress and strain components, we have

$$\begin{aligned} T_{11} &= T_1, & T_{23} &= T_{32} = T_4, \\ T_{22} &= T_2, & T_{31} &= T_{13} = T_5, \\ T_{33} &= T_3, & T_{12} &= T_{21} = T_6, \end{aligned} \tag{9}$$

and

$$\begin{aligned} S_{11} &= S_1, & 2S_{23} &= 2S_{32} = S_4, \\ S_{22} &= S_2, & 2S_{31} &= 2S_{13} = S_5, \\ S_{33} &= S_3, & 2S_{12} &= 2S_{21} = S_6. \end{aligned} \tag{10}$$

The constitutive relations in (3) take the following form:

$$T_p = c_{pq} S_q, \tag{11}$$

where $c_{pq} = c_{qp}$.

There are several variational formulations of the theory of linear elasticity. Two are particularly relevant to the present paper. One is the Hu-Washizu principle [3,4]. For an elastic body occupying a volume $v$, the functional of the Hu-Washizu principle is

$$\Pi(\mathbf{u}, \mathbf{S}, \mathbf{T}) = \int_v [U(\mathbf{S}) - T_{ij} S_{ij} + T_{ij} u_{i,j}] dv, \tag{12}$$

where boundary terms are unimportant to the purpose of this paper and are dropped for simplicity. The body force is also dropped for the same reason. The first variation of $\Pi$ is

$$\delta\Pi = \int_v \left[ \left( \frac{\partial U}{\partial S_{ij}} - T_{ij} \right) \delta S_{ij} + (u_{i,j} - S_{ij}) \delta T_{ij} - T_{ij,j} \delta u_i \right] dv. \tag{13}$$

Hence, the stationary condition of $\Pi$ is

$$\begin{aligned} T_{ji,j} &= 0, \\ T_{ij} &= \frac{\partial U}{\partial S_{ij}}, \\ S_{ij} &= \frac{u_{i,j} + u_{j,i}}{2}. \end{aligned} \tag{14}$$

The other relevant variational principle is the Hellinger-Reissner principle [5,6] whose functional, first variation, and stationary condition are

$$\Pi^*(\mathbf{u}, \mathbf{T}) = \int_v [-U^*(\mathbf{T}) + T_{ij} u_{i,j}] dv, \tag{15}$$

$$\delta\Pi^* = \int_v \left[ \left( -\frac{\partial U^*}{\partial T_{ij}} + u_{i,j} \right) \delta T_{ij} - T_{ij,j} \delta u_i \right] dv, \tag{16}$$

$$\begin{aligned} T_{ji,j} &= 0, \\ \frac{u_{i,j} + u_{j,i}}{2} &= \frac{\partial U^*}{\partial T_{ij}}. \end{aligned} \tag{17}$$

## 3. A Partial Complementary Energy Density, its Variational Principle and Application in Elastic Plates

In this section we introduce a partial complementary energy density, establish its variational principle, and show that they are useful in elastic plates. We use monoclinic crystals as an example. Their stiffness matrix is [2]:

$$[c_{pq}] = \begin{pmatrix} c_{11} & c_{12} & c_{13} & c_{14} & 0 & 0 \\ c_{21} & c_{22} & c_{23} & c_{24} & 0 & 0 \\ c_{31} & c_{32} & c_{33} & c_{34} & 0 & 0 \\ c_{41} & c_{42} & c_{43} & c_{44} & 0 & 0 \\ 0 & 0 & 0 & 0 & c_{55} & c_{56} \\ 0 & 0 & 0 & 0 & c_{65} & c_{66} \end{pmatrix}. \tag{18}$$

The constitutive relations corresponding to (18) are

$$\begin{aligned}
T_1 &= c_{11}S_1 + c_{12}S_2 + c_{13}S_3 + c_{14}S_4, \\
T_2 &= c_{12}S_1 + c_{22}S_2 + c_{23}S_3 + c_{24}S_4, \\
T_3 &= c_{13}S_1 + c_{23}S_2 + c_{33}S_3 + c_{34}S_4, \\
T_4 &= c_{14}S_1 + c_{24}S_2 + c_{34}S_3 + c_{44}S_4, \\
T_5 &= c_{55}S_5 + c_{56}S_6, \\
T_6 &= c_{56}S_5 + c_{66}S_6.
\end{aligned} \quad (19)$$

We separate (19) into two groups consisting of (19)$_{1,3,5}$ and (19)$_{2,4,6}$, respectively:

$$\begin{aligned}
T_1 &= c_{11}S_1 + c_{12}S_2 + c_{13}S_3 + c_{14}S_4, \\
T_3 &= c_{13}S_1 + c_{23}S_2 + c_{33}S_3 + c_{34}S_4, \\
T_5 &= c_{55}S_5 + c_{56}S_6,
\end{aligned} \quad (20)$$

$$\begin{aligned}
T_2 &= c_{12}S_1 + c_{22}S_2 + c_{23}S_3 + c_{24}S_4, \\
T_4 &= c_{14}S_1 + c_{24}S_2 + c_{34}S_3 + c_{44}S_4, \\
T_6 &= c_{56}S_5 + c_{66}S_6.
\end{aligned} \quad (21)$$

We then solve (21) for expressions of $S_2$, $S_4$ and $S_6$ in terms of $S_1$, $S_3$, $S_5$, $T_2$, $T_4$ and $T_6$, and substitute these expressions into (20) for expressions of $T_1$, $T_3$ and $T_5$ in terms of $S_1$, $S_3$, $S_5$, $T_2$, $T_4$ and $T_6$. The results can be arranged into

$$\begin{Bmatrix} T_1 \\ -S_2 \\ T_3 \\ -S_4 \\ T_5 \\ -S_6 \end{Bmatrix} = \begin{pmatrix} \gamma_{11} & \gamma_{12} & \gamma_{13} & \gamma_{14} & 0 & 0 \\ \gamma_{12} & \gamma_{22} & \gamma_{23} & \gamma_{24} & 0 & 0 \\ \gamma_{13} & \gamma_{23} & \gamma_{33} & \gamma_{34} & 0 & 0 \\ \gamma_{14} & \gamma_{24} & \gamma_{34} & \gamma_{44} & 0 & 0 \\ 0 & 0 & 0 & 0 & \gamma_{55} & \gamma_{56} \\ 0 & 0 & 0 & 0 & \gamma_{56} & \gamma_{66} \end{pmatrix} \begin{Bmatrix} S_1 \\ T_2 \\ S_3 \\ T_4 \\ S_5 \\ T_6 \end{Bmatrix}, \quad (22)$$

where

$$\begin{aligned}
\gamma_{11} &= c_{11} - \frac{c_{12}^2 c_{44} + c_{14}^2 c_{22} - 2c_{12}c_{24}c_{14}}{\Delta}, \quad \gamma_{12} = \frac{c_{12}c_{44} - c_{14}c_{24}}{\Delta} \\
\gamma_{13} &= c_{13} - \frac{c_{23}c_{12}c_{44} + c_{34}c_{14}c_{22} - c_{23}c_{14}c_{24} - c_{34}c_{12}c_{24}}{\Delta}, \quad \gamma_{14} = \frac{c_{14}c_{22} - c_{12}c_{24}}{\Delta}, \\
\gamma_{22} &= -\frac{c_{44}}{\Delta}, \quad \gamma_{23} = \frac{c_{44}c_{23} - c_{43}c_{24}}{\Delta}, \quad \gamma_{24} = \frac{c_{24}}{\Delta}, \\
\gamma_{33} &= c_{33} - \frac{c_{23}^2 c_{44} + c_{34}^2 c_{22} - 2c_{23}c_{34}c_{24}}{\Delta}, \quad \gamma_{34} = \frac{c_{34}c_{22} - c_{32}c_{24}}{\Delta}, \\
\gamma_{44} &= -\frac{c_{22}}{\Delta}, \quad \gamma_{55} = c_{55} - \frac{c_{56}^2}{c_{66}}, \quad \gamma_{56} = \frac{c_{56}}{c_{66}}, \quad \gamma_{66} = -\frac{1}{c_{66}}, \\
\Delta &= c_{22}c_{44} - c_{24}^2.
\end{aligned} \quad (23)$$

(22) represents the constitutive relations of monoclinic crystals in a form different from and equivalent to (19). We write (22) in a more compact form as

$$Y_p = \gamma_{pq} X_q, \quad (24)$$

where $p$ and $q$ range from 1 to 6, and

$$\begin{aligned} \mathbf{X} &= \{S_1 \quad T_2 \quad S_3 \quad T_4 \quad S_5 \quad T_6\}, \\ \mathbf{Y} &= \{T_1 \quad -S_2 \quad T_3 \quad -S_4 \quad T_5 \quad -S_6\}. \end{aligned} \qquad (25)$$

If we define a partial complementary energy density by

$$\tilde{U}(\mathbf{X}) = \frac{1}{2} \gamma_{pq} X_p X_q, \qquad (26)$$

then (24) can be obtained from

$$Y_p = \frac{\partial \tilde{U}(\mathbf{X})}{\partial X_p}. \qquad (27)$$

We note that the above procedure of introducing $\tilde{U}(\mathbf{X})$ indicates that $\tilde{U}(\mathbf{X})$ is related to $U(\mathbf{X})$ through the following partial Legendre transform:

$$\tilde{U}(\mathbf{X}) = U(\mathbf{S}) - T_2 S_2 - T_4 S_4 - T_6 S_6, \qquad (28)$$

in which $S_2$, $S_4$ and $S_6$ are viewed as functions of $S_1$, $S_3$, $S_5$, $T_2$, $T_4$ and $T_6$.

An example of the direct application of $\tilde{U}(\mathbf{X})$ is the constitutive relations for the extension of thin plates. Consider a plate of monoclinic crystals bounded by two planes at $x_2 = \pm h$ with $x_2$ as the plate normal. For the extension of such a plate, the main stress components are the in-plane ones, $T_{11}$, $T_{33}$ and $T_{13}$. In order to develop a two-dimensional theory for the extension of the plate, the stress relaxation condition of setting the relatively small stress components $T_{22} = T_{23} = T_{21} = 0$ or $T_2 = T_4 = T_6 = 0$ [2,7] needs to be applied to the plate constitutive relations. The stress relaxation is normally done by solving (21) with $T_2 = T_4 = T_6 = 0$ for expressions of $S_2$, $S_4$ and $S_6$ in terms of the in-plane strains $S_1$, $S_3$, $S_5$ and then substituting the expressions into (20) for expressions of $T_1$, $T_3$ and $T_5$ in terms of $S_1$, $S_3$, $S_5$. This involves some algebra when the usual constitutive relations in (3) or (7) from the strain energy density in (2) or the complementary energy density in (6) are used. However, if $\tilde{U}(\mathbf{X})$ and the corresponding constitutive relations in (22) are used, the stress relaxation can be directly imposed by setting $T_2 = T_4 = T_6 = 0$ in (22). Then (22)$_{1,3,5}$ directly give the relaxed constitutive relations for the extension of thin plates, and (22)$_{2,4,6}$ can be used to calculate $S_2$, $S_4$ and $S_6$ once the in-plane strains $S_1$, $S_3$, $S_5$ have been determined from solving the plate extensional problem.

For convenience we introduce a new index convention that $mn$=22, 23 or 32, and 21 or 12. $ab$=11, 33, and 13 or 31. From the Hu-Washizu functional in (12) we construct a new functional based on $\tilde{U}(\mathbf{X})$:

$$\tilde{\Pi} = \int_v [U(\mathbf{S}) - T_{mn}S_{mn} - T_{ab}S_{ab} + T_{ij}u_{i,j}]dv$$
$$= \int_v [\tilde{U}(\mathbf{X}) - T_{ab}S_{ab} + T_{ij}u_{i,j}]dv \qquad (29)$$
$$= \tilde{\Pi}(\mathbf{u}, \mathbf{X}, Y_{ab}).$$

It can be verified that the first variation of $\tilde{\Pi}$ is

$$\delta\tilde{\Pi} = \int_v \left[ \frac{\partial \tilde{U}}{\partial S_{ab}}\delta S_{ab} + \frac{\partial \tilde{U}}{\partial T_{mn}}\delta T_{mn} - T_{ab}\delta S_{ab} - S_{ab}\delta T_{ab} + T_{ij}\delta u_{i,j} + u_{i,j}\delta T_{ij} \right] dv$$
$$= \int_v \left[ \left(\frac{\partial \tilde{U}}{\partial S_{ab}} - T_{ab}\right)\delta S_{ab} + \left(\frac{\partial \tilde{U}}{\partial T_{mn}} + u_{m,n}\right)\delta T_{mn} + (u_{a,b} - S_{ab})\delta T_{ab} - T_{ij,j}\delta u_i \right] dv. \qquad (30)$$

Hence the stationary condition of $\tilde{\Pi}$ is

$$T_{ji,j} = 0,$$
$$T_{ab} = \frac{\partial \tilde{U}}{\partial S_{ab}}, \quad ab = 11, 33, \text{ and } 13 \text{ or } 31,$$
$$-\frac{u_{m,n} + u_{n,m}}{2} = \frac{\partial \tilde{U}}{\partial T_{mn}}, \quad mn = 22, 23 \text{ or } 32, \text{ and } 21 \text{ or } 12, \qquad (31)$$
$$S_{ab} = \frac{u_{a,b} + u_{b,a}}{2}, \quad ab = 11, 33, \text{ and } 13 \text{ or } 31.$$

Comparing (31) with (14) and (17), we see that the variational principle based on $\tilde{U}$ and $\tilde{\Pi}$ is somewhere between the Hu-Washizu principle and the Hellinger-Reissner principle.

For the application of $\tilde{\Pi}$, we consider the derivation of the two-dimensional equations for the extension of a thin plate of monoclinic crystals. We rewrite $\tilde{\Pi}$ as

$$\tilde{\Pi} = \int_v [\tilde{U}(\mathbf{X}) - T_{ab}S_{ab} + T_{ij}u_{i,j}]dv$$
$$= \int_v [\tilde{U}(\mathbf{X}) - T_{ab}S_{ab} + T_{ab}u_{a,b} + T_{mn}u_{m,n}]dv. \qquad (32)$$

We then use the strain-displacement relation for $S_{ab}$ and the stress relaxation condition of $T_{mn} = 0$ in (32), i.e.,

$$S_{ab} = \frac{u_{a,b} + u_{b,a}}{2}, \quad T_{mn} = 0. \qquad (33)$$

With (33), $\tilde{\Pi}$ reduces to

$$\tilde{\Pi}(S_{ab}) = \int_v \tilde{U}(S_{ab})dv. \qquad (34)$$

For extensional deformation of thin plates in the $(x_1, x_3)$ plane, the extensional displacement fields are approximated by [2]

$$u_1 \cong u_1(x_1, x_3), \quad u_3 \cong u_3(x_1, x_3). \qquad (35)$$

Then

$$\delta \tilde{\Pi} = \int_v \frac{\partial \tilde{U}}{\partial S_{ab}} \delta S_{ab} dv = \int_v \frac{\partial \tilde{U}}{\partial S_{ab}} \delta u_{a,b} dv = \int_v T_{ab} \delta u_{a,b} dv$$
$$= \int_A dA \left( \int_{-h}^{h} T_{ab} dx_2 \right) \delta u_{a,b} = \int_A N_{ab} \delta u_{a,b} dA = \int_A -N_{ab,b} \delta u_a dA, \tag{36}$$

where $A$ is an arbitrary area in the plane of the plate and we have introduced the plate extensional resultants by [2]

$$N_{ab} = \int_{-h}^{h} T_{ab} dx_2. \tag{37}$$

The stationary condition of (36) is

$$N_{ba,b} = 0, \quad ab = 11, 33, \text{ and } 13 \text{ or } 31, \tag{38}$$

which is the familiar equilibrium equation for plate extension.

## 4. Another Partial Complementary Energy Density, its Variational Principle and Application in Elastic Rods

In this section we introduce another partial complementary energy density, establish its variational principle, and show that they are useful in the development of the theory of elastic rods. We use polarized ceramics as an example. For ceramics poled along $x_3$, their stiffness matrix is:

$$\begin{pmatrix} s_{11} & s_{12} & s_{13} & 0 & 0 & 0 \\ s_{12} & s_{11} & s_{13} & 0 & 0 & 0 \\ s_{13} & s_{13} & s_{33} & 0 & 0 & 0 \\ 0 & 0 & 0 & s_{44} & 0 & 0 \\ 0 & 0 & 0 & 0 & s_{44} & 0 \\ 0 & 0 & 0 & 0 & 0 & s_{66} \end{pmatrix}, \tag{39}$$

where $s_{66} = 2(s_{11} - s_{12})$. The constitutive relations corresponding to (39) are

$$\begin{aligned} S_1 &= s_{11} T_1 + s_{12} T_2 + s_{13} T_3, \\ S_2 &= s_{12} T_1 + s_{11} T_2 + s_{13} T_3, \\ S_3 &= s_{13} T_1 + s_{13} T_2 + s_{33} T_3, \\ S_4 &= s_{44} T_4, \\ S_5 &= s_{44} T_5, \\ S_6 &= s_{66} T_6. \end{aligned} \tag{40}$$

We solve (40)$_3$ for $T_3$:

$$T_3 = \frac{1}{s_{33}} (S_3 - s_{13} T_1 - s_{13} T_2). \tag{41}$$

We then substitute (41) into the right-hand sides of (40)$_{1,2}$ and rearrange the resulting (40)$_{1,2,4-6}$ and (41) in the following form:

$$\begin{Bmatrix} S_1 \\ S_2 \\ -T_3 \\ S_4 \\ S_5 \\ S_6 \end{Bmatrix} = \begin{pmatrix} s_{11}-\dfrac{s_{13}^2}{s_{33}} & s_{12}-\dfrac{s_{13}^2}{s_{33}} & \dfrac{s_{13}}{s_{33}} & 0 & 0 & 0 \\ s_{12}-\dfrac{s_{13}^2}{s_{33}} & s_{11}-\dfrac{s_{13}^2}{s_{33}} & \dfrac{s_{13}}{s_{33}} & 0 & 0 & 0 \\ \dfrac{s_{13}}{s_{33}} & \dfrac{s_{13}}{s_{33}} & -\dfrac{1}{s_{33}} & 0 & 0 & 0 \\ 0 & 0 & 0 & s_{44} & 0 & 0 \\ 0 & 0 & 0 & 0 & s_{44} & 0 \\ 0 & 0 & 0 & 0 & 0 & s_{66} \end{pmatrix} \begin{Bmatrix} T_1 \\ T_2 \\ S_3 \\ T_4 \\ T_5 \\ T_6 \end{Bmatrix}, \qquad (42)$$

which can be further written as

$$Y_p = \beta_{pq} X_q, \qquad (43)$$

where

$$\begin{aligned} \mathbf{X} &= \{T_1 \quad T_2 \quad S_3 \quad T_4 \quad T_5 \quad T_6\}^T, \\ \mathbf{Y} &= \{S_1 \quad S_2 \quad -T_3 \quad S_4 \quad S_5 \quad S_6\}^T. \end{aligned} \qquad (44)$$

If we define a partial complementary energy density by

$$\hat{U}(\mathbf{X}) = \frac{1}{2}\beta_{pq} X_p X_q, \qquad (45)$$

then (43) can be obtained from

$$Y_p = \frac{\partial \hat{U}(\mathbf{X})}{\partial X_p} = \beta_{pq} X_q. \qquad (46)$$

$\hat{U}(\mathbf{X})$ can be obtained from the strain energy density $U(\mathbf{S})$ or the complementary stress energy density $U^*(\mathbf{T})$ through the following partial Legendre transform:

$$\begin{aligned} \hat{U} &= U^* - T_3 S_3 \\ &= T_1 S_1 + T_2 S_2 + T_4 S_4 + T_5 S_5 + T_6 S_6 - U, \end{aligned} \qquad (47)$$

where (8) has been used.

A simple example of the application of $\hat{U}(\mathbf{X})$ is the constitutive relations for the extension of thin rods. Consider a rod of ceramics poled along the axial direction $x_3$. For the extensional deformation of the rod, the main stress component is $T_3$. The stress relaxation condition of setting the small stress components $T_1 = T_2 = T_4 = T_5 = T_6 = 0$ [8] is needed to develop a one-dimensional theory. The stress relaxation can done using (40) easily for polarized ceramics but for materials with stronger anisotropy the algebra may be significant. However, if $\hat{U}(\mathbf{X})$ and the corresponding constitutive relations in (42) are used, the stress relaxation can be directly imposed on (42) by setting $T_1 = T_2 = T_4 = T_5 = T_6 = 0$. Then $(42)_3$ directly gives the relaxed constitutive relation for the extension of thin rods, i.e., $T_3 = S_3 / s_{33}$.
At the same time, $(42)_{1,2}$ can be used to calculate the lateral strains $S_1$ and $S_2$ due to the axial extension through Poisson's effect.

From the Wu-Washizu functional in (12) we construct a new functional based on $\hat{U}(\mathbf{X})$:

$$\hat{\Pi} = \int_v [U(\mathbf{S}) - T_{ij}S_{ij}|_{ij \neq 33} - T_{33}S_{33} + T_{ij}u_{i,j}]dv$$
$$= \int_v [-\hat{U}(\mathbf{X}) - T_{33}S_{33} + T_{ij}u_{i,j}]dv \qquad (48)$$
$$= \hat{\Pi}(\mathbf{u}, \mathbf{X}, T_{33}).$$

The first variation of $\hat{\Pi}$ is

$$\delta\hat{\Pi} = \int_v \left[ -\frac{\partial \hat{U}}{\partial T_{ij}}\delta T_{ij} \bigg|_{ij \neq 33} - \frac{\partial \hat{U}}{\partial S_{33}}\delta S_{33} - T_{33}\delta S_{33} - S_{33}\delta T_{33} + u_{i,j}\delta T_{ij}|_{ij \neq 33} + u_{3,3}\delta T_{33} + T_{ij}\delta u_{i,j} \right]dv$$
$$= \int_v \left[ \left(-\frac{\partial \hat{U}}{\partial T_{ij}} + u_{i,j}\right)\delta T_{ij} \bigg|_{ij \neq 33} - \left(\frac{\partial \hat{U}}{\partial S_{33}} + T_{33}\right)\delta S_{33} + (u_{3,3} - S_{33})\delta T_{33} - T_{ij,j}\delta u_i \right]dv.$$
$$(49)$$

The stationary condition of $\hat{\Pi}$ is

$$T_{ji,j} = 0,$$
$$-T_{33} = \frac{\partial \hat{U}}{\partial S_{33}},$$
$$\frac{u_{i,j} + u_{j,i}}{2} = \frac{\partial \hat{U}}{\partial T_{ij}}, \quad ij \neq 33, \qquad (50)$$
$$S_{33} = u_{3,3}.$$

For the application of $\hat{\Pi}$, we consider the derivation of the one-dimensional equations for the extension of a thin ceramic rod poled in the axial direction $x_3$. We rewrite $\hat{\Pi}$ as

$$\hat{\Pi} = \int_v [-\hat{U}(\mathbf{X}) - T_{33}S_{33} + T_{33}u_{3,3} + T_{ij}u_{i,j}|_{ij \neq 33}]dv. \qquad (51)$$

We then use the strain-displacement relation of $S_{33}$ and the stress relaxation of $T_{ij} = 0$ when $ij \neq 33$ in (51), i.e.,

$$S_{33} = u_{3,3}, \quad T_1 = T_2 = T_4 = T_5 = T_6 = 0. \qquad (52)$$

With (52), $\hat{\Pi}$ reduces to

$$\hat{\Pi}(S_{33}) = \int_v \hat{U}(S_{33})dv. \qquad (53)$$

For the extensional deformation of thin rods, the relevant displacement component is

$$u_3 \cong u_3(x_3). \qquad (54)$$

Then, for a rod of length $L$ and cross sectional area $A$,

$$\delta \hat{\Pi} = \int_v \frac{\partial \hat{U}}{\partial S_{33}} \delta S_{33} dv = \int_v -T_{33} \delta u_{3,3} dv = \int_L dx_3 \int_A (-T_{33} \delta u_{3,3}) dA$$
$$= \int_L dx_3 \left( \int_A T_{33} dA \right)(-\delta u_{3,3}) = \int_L N_{33}(-\delta u_{3,3}) dx_3 = \int_L N_{33,3} \delta u_3 dx_3, \quad (55)$$

where the plate extensional resultant is defined by

$$N_{33} = \int_A T_{33} dA. \quad (56)$$

The stationary condition of (55) is

$$N_{33,3} = 0, \quad (57)$$

which is the familiar equilibrium equation for rod extension.

## 5. Conclusions

Partial complementary energy densities with mixed strain and stress components may be introduced. Variational principles based on the partial complementary energy densities may be constructed accordingly. In addition to $\tilde{U}$ and $\hat{U}$, other energy densities and variational principles are possible and similar. These energy densities and variational principles are useful in the development of two- and one-dimensional theories of elastic plates and rods.